\documentclass[pre,aps,preprint,showpacs]{revtex4}
\usepackage{graphicx}
\usepackage{amsfonts,amssymb,amsmath,epsfig,amscd,ulem}

\def\bi{\begin{itemize}}

\def\ei{\end{itemize}}
\def\be{\begin{equation}}
\def\ee{\end{equation}}
\def\bea{\begin{eqnarray}}
\def\eea{\end{eqnarray}}

\def\gdot{\dot\gamma}

\DeclareTextSymbol{\degre}{OT1}{23}

\begin{document}

\author{G. Ovarlez$^{1}$, S. Rodts$^{1}$, A. Ragouilliaux$^{1}$, P. Coussot$^{1}$, J. Goyon$^{2}$, A. Colin$^{2}$}
\affiliation{$^{1}$Universit\'e Paris Est - Institut Navier\\ LMSGC (LCPC-ENPC-CNRS)\\
2, all\'ee Kepler, 77420 Champs-sur-Marne, France\\ $^{2}$LOF,
Universit\'e Bordeaux 1, UMR CNRS-Rhodia-Bordeaux 1 5258\\ 33608
Pessac cedex, France}

\title{Wide gap Couette flows of dense emulsions: local concentration measurements, and comparison between macroscopic and local constitutive law measurements through MRI}

\begin{abstract}
Flows of dense emulsions show many complex features among which
long range nonlocal effects pose problem for macroscopic
characterization. In order to get round this problem, we study the
flows of several dense emulsions, with droplet size ranging from
0.3 to 40$\mu$m, in a wide gap Couette geometry. We couple
macroscopic rheometric experiments and local velocity measurements
through MRI techniques. As concentration heterogeneities are
expected in the wide gap Couette flows of multiphase materials, we
also designed a new method to measure the local droplet
concentration in emulsions with a MRI device. In contrast with
dense suspensions of rigid particles where very fast migration
occurs under shear in wide gap Couette flows, we show for the
first time that no migration takes place in dense emulsions even
for strain as large as 100000 in our systems. As a result of the
absence of migration and of finite size effect, we are able to
determine very precisely the local rheological behavior of several
dense emulsions. As the materials are homogeneous, this behavior
can also be inferred from purely macroscopic measurements. We thus
suggest that properly analyzed purely macroscopic measurements in
a wide gap Couette geometry can be used as a tool to study the
local constitutive laws of dense emulsions. All behaviors are
basically consistent with Herschel-Bulkley laws of index 0.5. The
existence of a constitutive law accounting for all flows contrasts
with previous results obtained within a microchannel by Goyon {\it
et al.} [Nature {\bf 454}, 84 (2008)]: the use of a wide gap
Couette geometry is likely to prevent here from nonlocal finite
size effects; it also contrasts with the observations of B{\'e}cu
{\it et al.} [Phys. Rev. Lett. {\bf 96}, 138302 (2006)]. We also
evidence the existence of discrepancies between a perfect
Herschel-Bulkley behavior and the observed local behavior at the
approach of the yield stress due to slow shear flows below the
apparent yield stress in the case of a strongly adhesive emulsion.
\end{abstract}
 \maketitle

\section{Introduction}\label{section_introduction}

Emulsions are mixtures of two immiscible fluids consisting of
droplets of one phase dispersed into the other, stabilized against
coalescence by surfactant. At low droplets volume fraction, the
emulsions have basically a Newtonian behavior \cite{Larson1999}.
When the volume fraction of the dispersed phase is increased,
droplets come into contact. If a small stress is applied to a
dense emulsion, the interfaces between the droplets can be
strained to store surface energy; it results in an elastic
response \cite{Larson1999}. Above a yield stress, they flow as a
result of droplets rearrangements \cite{Larson1999}. The flow
behavior of dense emulsions can be measured in classical
rheological experiments, and seems to be well represented by a
Herschel-Bulkley model $\tau(\gdot)=\tau_c+\eta_{_{HB}}\gdot^n$;
in monodisperse emulsions, the exponent $n$ has been found to vary
between $2/3$ and $1/2$ for volume fractions varying between 0.58
and 0.65 \cite{Mason1996a}. At first glance, dense emulsions
behavior thus seems to be well understood and modelled.

However, it has been shown recently that the behavior of dense
emulsions \cite{Salmon2003,becu2006,Goyon2008}, and more generally
of pasty materials \cite{Coussot2005,moller2006}, may be much more
complex than what can be inferred from simple rheometric
experiments, particularly at the approach of the yield stress. The
reason is that the flow behavior of materials is usually studied
in viscosimetric flows \cite{Coussot2005}, in which one measures
macroscopic quantities (a torque $T(\Omega)$ vs. a rotational
velocity $\Omega$). A constitutive law relating the shear stress
$\tau(\gdot)$ to the shear rate $\gdot$ in any flow can then be
derived easily from these macroscopic measurements provided that
the flow is homogeneous. However, in complex fluids, this last
requirement may not be fulfilled as the flow may be localized, as
observed in many pasty materials
\cite{Coussot2002b,Raynaud2002,Coussot2005,Huang2005,Ragouilliaux2007,Ovarlez2006,becu2006},
even in the cone-and-plate geometry \cite{Coussot2002b}. Another
problem of importance, that can complicate the analysis of the
rheological data, is the existence of slippage of dense emulsions
at the walls \cite{Salmon2003,hollingsworth2004,Goyon2008} even
with slightly roughened walls \cite{Goyon2008}.

These problems can be got round by coupling macroscopic torque
measurements and local measurements of the shear rate, e.g.
through MRI techniques \cite{Raynaud2002}, optical methods
\cite{Salmon2003} or sound methods \cite{becu2006}, in order to
account for these heterogeneities and to get a proper constitutive
law. Local measurements in dense emulsions have recently yielded
surprising results \cite{Salmon2003,becu2006,Goyon2008}: in some
emulsions, it is impossible to find a single constitutive law
compatible with all flows. The constitutive law of the material
then seems to depend on the velocity at the inner cylinder in a
Couette cell \cite{Salmon2003,becu2006} or the pressure gradient
in a Poiseuille cell \cite{Goyon2008}. Three main reasons have
been evoked to explain the apparent absence of a constitutive law:
thixotropy, non local effects, and shear-induced migration of the
droplets.

Thixotropy may be a cause from the apparent absence of a
constitutive law. Long times may indeed be needed to reach a
steady state. When the measurements are not performed in this
steady state, it may then leave the idea that different laws are
required to describe the flow of the sample. Such long times to
reach steady state have been evidenced in some experiments dealing
with highly adhesive emulsion \cite{becu2006,
Ragouilliaux2006,Ragouilliaux2007}. Importantly, \citet{becu2006},
although they claim to have reached a steady state, were unable to
account for the flows of their adhesive emulsion with a single
constitutive law. Note that we will study again the emulsion of
\citet{becu2006} in this paper.

The apparent absence of a single local constitutive law may also
be the signature of non-local phenomena. Picard {\it et al.}
\cite{Picard2005} have actually proposed a non local model to
describe the flows of jammed materials (and thus dense emulsions),
based upon the cooperativity of the flow. In these systems, flow
occurs via a succession of elastic deformation and irreversible
plastic events. These localized plastic events induce a relaxation
of the stress on the whole system. Their zone of influence is
quantified by a length $\xi$. In presence of high stress gradient
or in the vicinity of a surface, this process creates non local
effects: indeed, in these situations, the rate of rearrangements
of the neighboring fluid differs from what would be the bulk
rearrangement rate. As a consequence, these effects affect
drastically the rheological behavior when $\xi$ is comparable to
the size of the confinement. Such behavior has been recently
pointed out by \citet{Goyon2008}. From experimental data, the
authors show that the cooperativity length $\xi$ is zero below the
jamming concentration $\phi_m$ and is typically a few oil droplet
diameters above $\phi_m$. Using a wide gap should then prevent us
from being sensitive to these non local effects. We will check
this feature by studying the flows of the \citet{Goyon2008}
emulsion in this paper.

Particle migration may also be a cause for the apparent lack of a
local constitutive law \cite{Ovarlez2006}. In multiphase
materials, wide gap Couette flows are indeed known to lead to
concentration heterogeneities due to shear-induced migration of
the dispersed elements toward the low shear zones. The migration
of deformable particles, and particularly drops, has been much
less studied than the one of rigid particles. Single droplets have
been found to migrate away from the rigid walls in any shear flow,
due to asymmetric flows around deformed droplets
\cite{karnis1967}. This leads to an equilibrium position somewhere
between the walls, exactly at the center of the gap in the case of
a narrow gap Couette geometry, nearer from the inner cylinder in
the case of a wide gap Couette geometry \cite{hollingsworth2006}.
When dealing with a dilute emulsion, it has been observed, in
contrast with dilute suspensions of rigid particles, that the
equilibrium distribution of droplets is parabolic around a
position that is near the equilibrium position of single droplets
\cite{king2000,hudson2003,hollingsworth2006}. The theoretical
explanation is that all particles would tend to accumulate at the
same equilibrium position between the walls, but their
distribution is broadened by their binary collisions (this second
process may be modelled as it is for rigid particles). Only rather
dilute emulsions (up to 10\% of droplets) have been studied so
far, and nothing is known about what happens for higher droplet
concentrations. In this case, one would expect the wall effect on
the droplet migration to be negligible compared to the effect of
the interactions between droplets; then, one would expect
migration to product the same effects as in suspensions, with the
same kinetics as it has basically the same physical origin. We can
thus expect that relevant information for migration in dense
emulsions can be found in the literature dealing with suspensions
of rigid particles. In suspensions of noncolloidal rigid
particles, migration has been observed in many situations:
wide-gap Couette flows
\cite{Leighton1987b,Phillips1992,Ovarlez2006,Graham1991,Corbett1995},
parallel-plate flows \cite{Barentin2004}, pipe flows
\cite{Sinton1991,Lyon1998}. In wide-gap Couette flows, the
consequence of migration is an excess of particles near the outer
cylinder. Migration is related to the shear-induced diffusion of
particles \cite{Leighton1987b,Phillips1992}: the gradients in
shear rate that exist in all but the cone and plate geometry
generate a particle flux towards the low shear zones (the outer
cylinder in the case of the Couette geometry), which is
counterbalanced by a particle flux due to viscosity gradients. In
an alternative model \cite{Nott1994,Mills1995}, particle fluxes
counterbalance the gradients in normal stresses. Most experiments
\cite{Graham1991,Phillips1992,Corbett1995} observe that migration
in suspensions of volume fraction up to 55\% is rather slow, in
accordance with its diffusive origin. However, it was recently
found in a Couette geometry that in the case of very dense
suspensions (up to 60\%), which exhibit an apparent yield stress,
migration is almost instantaneous \cite{Ovarlez2006} (it lasts for
a few revolutions) so that it may be unavoidable. A reason for
this very fast kinetics may be that particles are closely packed
together: any shear may then push the particles towards the outer
cylinder with an instantaneous long range effect on the particle
concentration. Migration may then be expected to be very fast for
dense emulsions exhibiting a yield stress, i.e. in which droplets
are in contact as the particles in the very dense suspensions of
\cite{Ovarlez2006}.

In this paper, we address the questions of the existence and the
determination of a local constitutive law for the flows of dense
emulsions. In this aim, we use a wide gap Couette cell to avoid
non local effects and we study in detail the question of
shear-induced droplet migration in these systems. We also take
care of performing our experiments in a steady state. We propose
to couple macroscopic measurements and local measurements of
concentration and velocity through MRI techniques during the wide
gap Couette flows of dense emulsions. A new method designed to
measure the local droplet concentration in emulsions is developed
and we seek for the occurrence of migration on many formulations.
Local measurements of the constitutive law are then compared to
purely macroscopic measurements.

In Sec.~\ref{section_display}, we present the materials and the
experimental setup. We present the concentration profiles obtained
on all materials after long time experiments in
Sec.~\ref{section_concentration}. The velocity profiles are shown
in Sec.~\ref{section_velocity}. The macroscopic rheometrical
measurements are displayed in Sec.~\ref{section_torque}, and a
purely macroscopic determination of the constitutive law is
presented. The results are analyzed in Sec.~\ref{section_local}:
we determine locally the constitutive law of the material from the
velocity profiles and we compare macroscopic and local
measurements of the constitutive law.

\section{Materials and methods}\label{section_display}

\subsection{Emulsions}
As the timescale for the occurrence of migration is {\it a priori}
very sensitive to the droplet size $R$ (if a diffusive process is
involved, it should scale as $1/R^2$), we study dense emulsions of
4 different sizes: 0.3$\mu$m, 1$\mu$m, 6.5$\mu$m, and 40$\mu$m.
Moreover, as problems in defining a local constitutive law seem to
be linked to the adhesion properties of the droplets
\cite{becu2006}, we took care of formulating adhesive and non
adhesive emulsions. Note in particular that we study the emulsions
of \citet{becu2006} and \citet{Goyon2008} which posed problems in
defining a single local constitutive law accounting for all their
flows.

The 0.3$\mu$m emulsion is the adhesive emulsion of
\citet{becu2006}; nearly monodisperse oil in water emulsions are
prepared by shearing a crude polydisperse emulsion, composed of
castor oil droplets in water stabilized by Sodium Dodecyl Sulfate
(SDS), within a narrow gap of 100 $\mu$m \cite{Mason1996b}. The
surfactant concentration within the aqueous phase is set to 8\%
wt. which leads to an adhesive emulsion \cite{becu2006}. The oil
volume fraction is 73\% and the droplet mean diameter is 0.3
$\mu$m with a polydispersity of about 20\% which is enough to
prevent crystallization.

The 1$\mu$m emulsion is a non adhesive emulsion. The preparation
of the sample is the same as the one used to get the adhesive
0.3$\mu$m emulsion. It is composed of silicone oil in water
stabilized by Sodium Dodecyl Sulfate (SDS). The surfactant
concentration within the aqueous phase is set to 8.5\% . The
viscosity of the oil is 1 Pa.s and the oil volume fraction is
75\%.

We prepared adhesive and non-adhesive oil in water emulsions, of
6.5 $\mu$m mean diameter with a polydispersity of about 20 $\%$.
These emulsions are composed of silicone droplets in a mixture of
glycerine and water (50\% w glycerine, 50\% w water) stabilized by
1\% wt of Brij and trimethyl tetradecyl ammonium bromide, sheared
in a narrow gap Couette cell. The trimethyl tetradecyl
concentration within the aqueous phase is set to 1 or 6.5 wt $\%$,
which leads to respectively adhesive and non-adhesive emulsions.
The oil viscosity is 1Pa.s. The oil volume fraction is 75\% in all
cases. Note that the non-adhesive 6.5$\mu$m emulsion is the same
as in \citet{Goyon2008}. Note that the use of glycerol in the
continuous phase prevents from obtaining accurate concentration
measurements with the method developed below. Only qualitative
results can be inferred from the measurements: in particular, if
the intensity profiles are unchanged after a flow, it means that
there is no migration.

We finally studied a large droplet brine in oil non-adhesive
emulsion, of 40 $\mu$m mean diameter with a polydispersity of
about 50 $\%$, at a 88\% concentration. The emulsion is prepared
by progressively adding the brine (solution of CaCl$_2$ at 300g/l
in water) in an oil-surfactant solution (HDF 2000 (Total Solvent)
+ Sorbitan monooleate at 2\%) under a controlled high shear with a
Silverson L4RT mixer.

\subsection{Rheological measurements}
We carried out experiments with a velocity controlled ''magnetic
resonance imaging (MRI) rheometer". Experiments are performed
within a wide-gap Couette geometry (inner cylinder radius
$R_i=4.1$cm, outer cylinder radius $R_o =6$cm, height $H=$11cm).
In order to avoid slip at the walls, sandpaper is glued on the
walls. Rheometric measurements are performed with a commercial
rheometer (Bohlin C-VOR 200) that imposes either the torque or the
rotational velocity (with a torque feedback). The whole geometry
is inserted in a 0.5T vertical MRI spectrometer (24/80 DBX by
Bruker). The MRI measurements allow us to get the local
orthoradial velocity and the local droplet concentration of the
material anywhere in the gap, with a radial resolution of 0.5 mm.
For details on the MRI sequence devoted to measure the local
velocity and its application to rheology, see
\cite{Rodts2004,Ovarlez2006}. A method devoted to obtain the local
droplet concentration was developed specifically for this study
and is presented hereafter.

\subsubsection*{Concentration measurements}

Proton NMR relies on the properties of hydrogen nuclei, which bear
a magnetic dipole. In a NMR experiment, the sample is put inside a
strong, static, and (as far as possible) homogeneous field $\vec
B_{\rm o}$. The magnetization of hydrogen nuclei then tend to
align along the static field. The sample develops a non zero
magnetization density $m_{\rm o}(\vec r)$ which follows the Curie
law: \bea m_{\rm o}(\vec r)=\rho_H(\vec r)B_{\rm
o}\frac{\gamma^2\hbar^2}{4k_BT}\label{eq_magnetization}\eea where
$\rho_H$ is the hydrogen density at position $\vec r$, $\gamma$ is
the gyromagnetic ratio of hydrogen nuclei, $\hbar$ is the Planck
constant, $k_B$ is the Boltzmann constant, and $T$ is the sample
temperature. $\rho_H$ can be easily computed for any sample from
its chemical formula and its density.

When put out of equilibrium, magnetization of liquids often
develops a threefold dynamic behavior combining a precession
motion around the permanent field of the magnet, and two
relaxation processes: a mono-exponential decay of transverse
components of $\vec m_{\rm o}(\vec r)$ with a characteristic time
$T_2$, and a mono-exponential relaxation of the longitudinal
component of $\vec m_{\rm o}(\vec r)$ towards its equilibrium
value, with a characteristic time $T_1$. Nuclear Magnetic
Resonance technique consists in series of magnetization
manipulation and measurements by means of externally applied
additional magnetic field, aiming at getting information about
physico-chemical properties of the sample. The Magnetic Resonance
Imaging extension of NMR permits to get, as far a possible, a
space resolved view of the local magnetization density, and can
make such analysis local and non-perturbative.

Because of signal to noise ratio limitations, the space resolution
of MRI facilities for imaging purposes can never be much better
than about one hundredth of the size of the experimental setup,
that is 1mm in our working conditions, which is much too big to
resolve independent droplets. Measuring the water/oil ratio in a
sample then requires to have a way to distinguish oil and water
magnetization inside a single pixel of a MRI image.

Up to now, three main discrimination routes were already reported
in emulsion literature. The first one relies on the so-called
chemical shift difference between water and oil. Actually, the
exact precession frequency around the permanent field is closely
related to the direct chemical neighborhood of hydrogen atoms.
Because all hydrogen nuclei of water are linked to oxygen atoms,
and most of hydrogens in oily material are linked either to carbon
or silicium atoms, water and oil magnetization always exhibit two
different precession frequency. This chemical shift is quite small
(about 4-4.5 ppm relative difference, whenever carbon or silicon
oils are used). Nevertheless this shift can be accurately observed
provided that $\vec B_{\rm o}$ inhomogeneities across the sample
have a much smaller amplitude than 4.5ppm of the mean $B_{\rm o}$
value (e.g., see spectrum by \citet{gotz2003}).Indeed, the
precession frequency is also proportional to local $\vec B_{\rm
o}$ intensity, and one should avoid confusion between the original
chemical shift and any spurious one originating from field
heterogeneities. Homogeneity constraints can be met when the fluid
is embedded in a container with a simple-shape. Reported studies
making use of the chemical shift contrast then often deal with
emulsions that are contained in a simple beaker, or even a
cylindrical pipe displayed parallel to $\vec B_{\rm o}$. Quite
early results deal with the use of chemical shift imaging (CSI)
for a quantitative assessment of oil/fat content in food emulsions
\cite{duce1994}, or the separation of oil and water components of
an emulsion during a filtration process \cite{yao1995}. More
recently, \citet{gotz2003} and \citet{hollingsworth2006} managed
to carry out quite well resolved local measurements of water/oil
concentrations and/or velocities in model suspensions flowing
through a pipe in a laminar regime. Unfortunately, the chemical
shift contrast can hardly be transposed in the case of a Couette
cell. The geometry is indeed more complex and involves edges
between materials with different magnetic susceptibilities, prone
to generate strong local field distortions. CSI applications to
Couette flow are thus more difficult and actually quite rare. For
instance, \citet{davila2003}, who studied the mixing process of
oil and water in an horizontal Couette like device reported quite
long measuring times and only poorly resolved pictures.

A second way to discriminate between water and oil in an emulsion
relies on $T_1$ relaxation time. $T_1$ (as well as $T_2$) indeed
strongly depend on a wide set of parameters, such as fluid
composition, diffusion coefficient, temperature, or the presence
of dissolved species. In water and oil, they usually turn out to
be quite different. For instance, in our working conditions (0.5T
magnetic field) $T_1$ and $T_2$ usually range between 2 and 3s in
bulk water (depending on water origin and dissolved minerals),
while many carbonated and silicon oils exhibit $T_1$ about few
100ms, and $T_2$ even smaller. Although the use of $T_2$ contrast
was already reported as a promising NMR tool of characterizing
droplet size in an emulsion \cite{johns2007}, previous works
involving Imaging possibilities of NMR prefer $T_1$ contrast. As
compared with chemical shift or $T_2$, $T_1$ physical value and
$T_1$ measurement techniques are indeed fairly insensitive to
field inhomogeneities. \citet{kauten1991} demonstrated this
contrast to give quantitative information on the creaming process
of an oil-water emulsion. More recently, few works showed its
ability to work in more severe conditions. \citet{marciani2004}
used $T_1$ to measure the fat fraction in an emulsion contained
inside the stomach of a human volunteer. \citet{hollingsworth2006}
also reported the use of a $T_1$ based method to follow the
droplet migration in a low concentrated emulsion under shearing in
a Couette cell.

A third way also used to study the creaming of an emulsion
\cite{newling1997}, consists in preparing a water-oil emulsion
based on deuterium oxide instead of normal water. Because proton
NMR is only sensitive to 1H nuclei, NMR or MRI measurements then
bring exclusive information on the oil phase.

In our work, we wanted to get independent information on oil and
water phase. Field inhomogeneities inside our Couette cell were
also in many places of the same order of magnitude as the chemical
shift. This was mainly due to the covering of the surfaces of the
cell with sand-paper, that was used so as to avoid slip problems.
$T_1$ contrast was then regarded as the best contrast source.
Although details of our measurement method was quite different
than that of \citet{hollingsworth2006} it was also based on the
so-called inversion-recovery technique. Starting from a
magnetization in equilibrium state, it consists in putting
magnetization upside down by means of an appropriate external
field, and then to let it relax towards its equilibrium value. In
a homogeneous sample with relaxation time $T_1$ and equilibrium
magnetization density $m_{\rm o}$, the magnetization about the
main magnetic field should then read \bea m(t)=m_{\rm
o}\Bigl(1-2e^{-t/T_1}\bigr)\eea where $t=0$ at the time of
inversion. To perform an accurate analysis, we found necessary to
consider additionally that the initial inversion can never be
perfect: the magnetization about the main magnetic field then
actually reads \bea m(t)=m_{\rm o}\Bigl(1-Ie^{-t/T_1}\bigr)\eea
where $I$ reflects the actual inversion conditions: $I$ would be
equal to 2 in perfect experimental conditions, but was actually
always found to be slightly less than 2. Its exact value and space
homogeneity depends on both the hardware and the sample.

In a water-oil emulsion, it is admitted that water and oil behave
as two independent phases: the resultant magnetization is then the
sum of the individual oil and water magnetization that relax
independently. The mean magnetization inside a water oil-emulsion
then reads \bea m(t)=\phi^w m_{\rm o}^w(1-I e^{-t/T_1^w})+\phi^o
m_{\rm o}^o(1-I e^{-t/T_1^o})\label{eq_relaxation} \eea where
$m_{\rm o}^w$ and $m_{\rm o}^o$ are the equilibrium magnetization
density of bulk water and oil respectively, $\phi^w$ and $\phi^o$
are the water and oil volume fractions ($\phi^w+\phi^o=1$), and
$T_1^w$ and $T_1^o$ are the longitudinal relaxation time of water
and oil respectively. $I$ is supposed to be the same in oil and
water.

From Eq.~\ref{eq_relaxation}, in order to observe only water
droplets in a water in oil emulsion, \citet{hollingsworth2006}
used the fact that there is always a time during this relaxation
process when the oil magnetization is zero. They made a MRI
picture of $m(t)$ at this time, based on the assumption that $I=2$
everywhere in the sample. However, we think that the accuracy of
such technique may be limited, because a wrong estimation of the
actual $I$ factor, if not exactly 2, or a wrong estimation of
relaxation times will always lead to a wrong estimation of the
time for which the oil magnetization is zero, and then, to an
incomplete oil cancelling.

Using in situ measurements of water and oil relaxation times, we
developed another approach which consists in performing 3 MRI
measurements of the magnetization relaxation, for 3 carefully
chosen waiting times, and then in solving an equation system,
letting $I$ as a free parameter. We then obtain space resolved
measurements of $I(\vec r)$, $\phi^w(\vec r)\,m_{\rm o}^w$ and
$\phi^o(\vec r)\,m_{\rm o}^o$ as averaged at the pixel scale. The
three waiting times are optimized for each sample in order to
minimize the uncertainty on the magnetization values. As compared
with the 'phase cancelling' technique, we calculated that this
method is much more accurate, and especially less sensitive to
$T_1$ uncertainties when $T_1$ values in water and oil are clearly
separated.

In order to get the material volume fraction from
Eq.~\ref{eq_relaxation}, in situ measurement of relaxation times
were found to be a necessary step of the experimental procedure.
Indeed, as far as relaxation phenomena are considered, although
water and oil mainly behave as if they would be independent
phases, they interact anyway together with the surfactant at the
surface of each droplet. Relaxation times in water may then be
influenced by the presence of the oil phase. Relaxation times in
oil may also differ from that of pure oil because aliphatic chains
of surfactant molecules mix with oil in the vicinity of the
droplet surface. The sensitivity to the droplet size was reported
to be quite dramatic for $T_2$ value in water/oil emulsion
containing a large amount of additional impurities
\cite{johns2007}. We believe in such system that $T_1$ and $T_2$
should qualitatively behave the same. In our samples, which are
made of quite pure water and oil phases, the $T_1$ difference
between the pure phases and the droplets was found to be of order
10\%. This difference is probably too small to provide a
quantitative sensitivity of our measurements to droplet size
changes in our systems. Anyway, $T_1$ changes, as small as they
may be, have to be taken into account for the accuracy of local
fraction measurements. Eventually, $T_1$ may also evolve with any
small change in the temperature of the sample. That is why before
each concentration measurement, the actual relaxation times were
measured.

As $T_1$ is sensitive to the details of the local structure of the
material, $T_1$ measurements should be ideally space resolved.
However, we worked here another way: we used an average $T_1$
measurement in the whole gap of the cell, discriminating water and
oil on the basis of their chemical shift (field inhomogeneities
were too strong for a CSI-based imaging, but were still OK for
$T_1$ measurements). The average $T_1$ values for oil and water
were then directly used to process the MRI data. Granted that
processed oil and water volume fractions were then always found to
be homogeneous within the gap of the cell, we had then an a
posteriori confirmation that our global $T_1$ measurement were
meaningful. One should be aware anyway that if any heterogeneity
would have been detected in measured volume fractions, then it
would have been necessary to shift to local $T_1$ measurements to
ensure very accurate quantitative results; measurements based on
the global $T_1$ measurement would nevertheless provide a relevant
estimation of the volume fractions.

Finally, in order to obtain the local volume fraction
$\phi^o\,(\vec r)$ and $\phi^w\,(\vec r)$ of each phases, it is
sufficient to remember that from Eq.~\ref{eq_magnetization}:
$m_{\rm o}^w= \rho_H^w(B_{\rm o}\gamma^2\hbar^2/4k_BT)$ and
$m_{\rm o}^o=\rho_H^o(B_{\rm o}\gamma^2\hbar^2/4k_BT)$. The
absolute uncertainty on the concentration measurement performed
with this method was found to be of 0.3\% from measurements
performed on a homogeneous emulsion. Note that when the proton
density of the oil phase is not known, the comparison of the
$\phi^w(\vec r)\,m_{\rm o}^w$ and $\phi^o(\vec r)\,m_{\rm o}^o$
measurements performed on the homogeneous emulsion (of known
volume fraction) allows to recover its value. Note also that a
simple way to check for changes in the local concentration, and to
evaluate the amplitude of these changes, consists in comparing the
water and oil intensity profiles within the gap after shear to the
profiles measured on the emulsion just after loading. These last
profiles are performed on a homogeneous emulsion, any change in
these profiles induced by shear would then mean that there is
migration.

\subsubsection*{Procedures}

In the experiments presented hereafter, we control the rotational
velocity of the inner cylinder. For all materials, we first apply
a constant rotational velocity of 20rpm to 100rpm until 10000 to
90000 revolutions have been completed. We record the evolution of
the velocity profiles and the concentration profiles in time in
order to check for the occurrence of migration (see
Sec.~\ref{section_concentration}). Afterwards, we measure the
stationary velocity profiles and the stationary torque exerted by
the material on the inner cylinder for various velocities ranging
from 0.1rpm to 100rpm (see Sec.~\ref{section_velocity} and
Sec.~\ref{section_torque}).

\section{Results} \label{section_results}

\subsection{Concentration profiles}\label{section_concentration}

In order to search for the occurrence of shear-induced migration
in dense emulsions, we sheared all materials for long times. We
chose to apply velocities ranging between 20 and 100 rpm; this
ensures that in all cases the whole gap is sheared (see
Sec.~\ref{section_velocity}). We also checked on the 6.5$\mu$m
emulsions that the same results are obtained when the flow is
localized during the experiment (i.e. when the material is sheared
at 1 rpm). The 0.3$\mu$m emulsion was sheared for 10000 revs, the
1$\mu$m emulsion was sheared for 90000 revs, both 6.5$\mu$m
emulsions were sheared for 35000 revs, and the 40$\mu$m emulsion
was sheared for 20000 revs. The corresponding local strains,
extracted from the velocity profiles (see
Sec.~\ref{section_velocity}) range between 5 times the number of
revolutions near the inner cylinder, and 0.25 times near the outer
cylinder.

\begin{figure}[htbp]
\includegraphics[width=9cm]{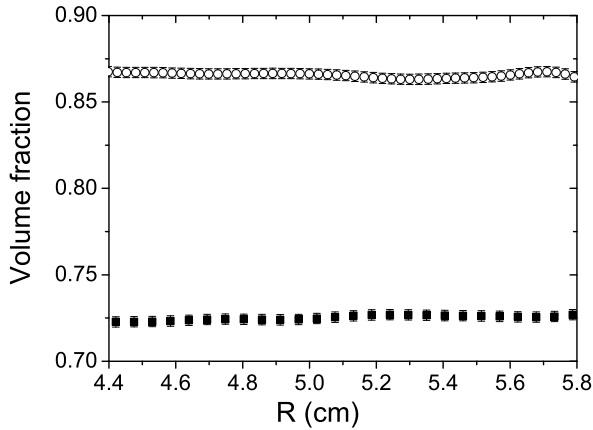}
\caption{Concentration profiles observed within the gap of a
Couette geometry after shearing a 0.3$\mu$m adhesive emulsion for
10000 revs (squares), and a 40$\mu$m non-adhesive emulsion for
20000 revs (open circles).} \label{fig_emul_concentration}
\end{figure}

In all the experiments, within the experimental uncertainty, we
observe that the materials remain homogeneous after shear: there
is no observable shear-induced migration. Results for the
0.3$\mu$m and the 40$\mu$m emulsion are depicted in
Fig.~\ref{fig_emul_concentration}. Note also that the relaxation
times $T_1$ of the oil and water phases were not observed to
change during all the experiments: as pointed out in
Sec.~\ref{section_display}, this is consistent with the fact that
the droplet size is unchanged by shear and that all the
measurements deal with the same homogeneous suspension.
Consistently, optical observations of the droplets after shear did
not reveal size changes.

As noted in Sec.~\ref{section_introduction}, up to now only rather
dilute emulsions have been studied: we cannot compare our results
with any result from the literature dealing with emulsions.
However, the relevant mechanism involved in migration of droplets
in dense emulsions should be basically the same as in suspensions
of rigid particles, namely a shear-induced diffusion of particles.
From this model and results from the literature of migration in
dense suspension, we can thus infer whether or not migration would
have been expected in our systems. In the diffusive model of
migration \cite{Leighton1987b,Phillips1992}, the particles undergo
a shear-induced diffusion characterized by a self-diffusion
coefficient $D=\bar D(\phi)\dot\gamma a^2$
\cite{Leighton1987a,Acrivos1995}, where $\phi$ is the particle
volume fraction, $\dot\gamma$ the shear rate, $a$ the particle
size, and $\bar D$ is a dimensionless coefficient whose dependence
on $\phi$ may theoretically be $\bar D(\phi)\propto\phi^2$ for
small $\phi$. The gradients in shear rate that exist in a Couette
geometry then generate a particle flux towards the outer cylinder,
which is counterbalanced by a particle flux due to viscosity
gradients. As qualitatively confirmed experimentally
\cite{Abbott1991,Corbett1995}, one would then expect the migration
phenomenon to last for a number of revolutions $
N_{migr}\propto(R_o-R_i)^3/(\bar{R} a^2\phi^2)$ until a stationary
heterogeneous profile is established, where $R_o$ and $R_i$ are
respectively the outer and inner radius, and
$\bar{R}=(R_o+R_i)/2$. Within the frame of this model, we can now
evaluate the expected $N_{migr}$ for complete migration in our
experiment from experimental results from literature. Actually, as
shown by \citet{Ovarlez2006} two inconsistent set of data exist,
depending on the concentration: migration has been shown to occur
much more rapidly for concentrations above 55\%. E.g, for
moderately dense suspensions, \citet{Corbett1995} find
$N_{migr}=2000$ revs for 140$\mu$m particles, at $\bar{\phi}=0.4$,
with $R_o=1.9$cm and $R_i=0.95$cm. In the case of our materials,
this would imply a value of $N_{migr}\approx 2.8.10^8$ revs for
the 0.3$\mu$m emulsion, $N_{migr}\approx 2.5.10^7$ revs for the
1$\mu$m emulsion, $N_{migr}\approx 6.10^5$ revs for the 6.5$\mu$m
emulsions, and $N_{migr}\approx 10000$ revs for the 40$\mu$m
emulsion. On the other hand for very dense suspensions,
\citet{Ovarlez2006} found that migration occurs during the first
50 revs within the same Couette geometry as our. This would imply
an expected value of $N_{migr}\approx 2.8.10^7$ revs for the
0.3$\mu$m emulsion, $N_{migr}\approx 2.5.10^6$ revs for the
1$\mu$m emulsion, $N_{migr}\approx 60000$ revs for the 6.5$\mu$m
emulsions, and $N_{migr}\approx 1000$ revs for the 40$\mu$m
emulsion in our experiments. Let us recall that $N_{migr}$ is the
number of revolutions expected for {\it complete} migration, and
that heterogeneities should be observable for a value of
$N_{migr}/10$ revolutions (see \citet{Abbott1991} and
\citet{Tetlow1998} for measurements of transient concentration
profiles in suspensions).

If a shear-induced migration mechanism similar to the one observed
in suspensions of rigid particles was acting in dense emulsions,
we should therefore have observed migration in the 40$\mu$m
emulsion and in the 6.5$\mu$m emulsions. On the other hand, our
experiments cannot be used to conclude directly for the failure of
a diffusive model of migration in the case of the smaller droplets
(0.3 and 1$\mu$m), but we can expect to predict what happens in
this case from the $a^2$ scaling of the diffusion process if this
process is relevant. Even if it impossible to conclude definitely
for the absence of migration for any strain, we can conclude that
if a diffusive process acts in dense emulsions, it is much slower
than the process involved in dense suspensions of rigid particles.
Importantly, our results provide reliable lower bounds for the
occurrence of migration in dense emulsions. For practical
purposes, in most flows and particularly in most rheometrical
studies of dense emulsions, it can therefore be claimed that the
emulsions remain homogeneous.

Surprisingly, our results contrast with what was observed by
\citet{Ovarlez2006} in dense suspensions of hard spheres where
very fast migration was found to occur. A first reason could be
that jamming prevents from migration in dense emulsions. However,
it was found that migration in dense suspensions can lead to
regions having a concentration higher than the maximum packing
fraction $\phi_m$ above which there is no more flow
\cite{Ovarlez2006}: this shows that jamming does not necessarily
prevents from migration. Another reason could be that
deformability of the particles play a central role. The shear
stresses involved in the flow are of the order of 100 to 200 Pa in
all materials, whereas the typical stress due to surface tension
range between 1000Pa for the 40$\mu$m droplets and 120000Pa for
the 0.3$\mu$m particles. The droplets may then be poorly
deformable (except for the 40$\mu$m emulsion) under shear, but
they must be deformed due to confinement and to their high
concentration. In particular they should have flat contacts which
may allow the droplets to slip easily past each other; this is a
major difference with the suspensions of hard spheres in which any
shear may push the particles towards the outer cylinder with an
instantaneous long range effect on the particle concentration
thanks to the force chains originating from direct frictional
contact forces.

Note that consistency of our observations with the modelling of
shear-induced migration based on normal stresses is difficult to
check as normal stresses in dense emulsions are poorly known. It
should be noted however that the normal stresses due to surface
tension should be of the order of the yield stress
\cite{Larson1999} in dense emulsions, while normal stresses tend
to diverge at the approach of the maximum packing fraction in
dense suspensions of rigid particles (this difference is linked to the
deformability of the droplets leading to flat contacts as pointed out above).
It is thus not unexpected
from this point of view that the kinetics of shear induced
migration in a Couette flow, which depends on the spatial
variation of the first normal stress difference and of the radial
normal stress \cite{Morris1999}, is very fast near the maximum
packing fraction in suspensions of hard spheres while it would
have to remain slow in dense emulsions.  Anyway, it is not
possible to be more quantitative at this stage and to state if
migration would have been expected to occur within this picture in
our experimental conditions.

\subsection{Velocity profiles}\label{section_velocity}

In these experiments, after a 5 min preshear at 100 rpm, we
measured the velocity profiles at various rotational velocities.
In all systems, the stationary velocity profiles were found to
develop within a few seconds and to remain stable for hours. This
means that these systems are not thixotropic \cite{Coussot2005}.
In the case of the 0.3 $\mu$m adhesive emulsion, this contrasts
with the long time evolution of the velocity profiles observed by
\citet{becu2006}; we will comment on this point below. The
stability of the velocity profiles is also consistent with the
absence of shear-induced droplet migration in all systems: as the
behavior of dense emulsions depends on their concentration,
changes in the concentration profiles in time would have led to
changes in the velocity profiles in time. This feature is
important: in the absence of a method to measure the concentration
profiles, the stability of the velocity profiles can be used as a
proof of the absence of migration as long as the migration process
is not nearly instantaneous.

\begin{figure}[htbp]
\includegraphics[width=8cm]{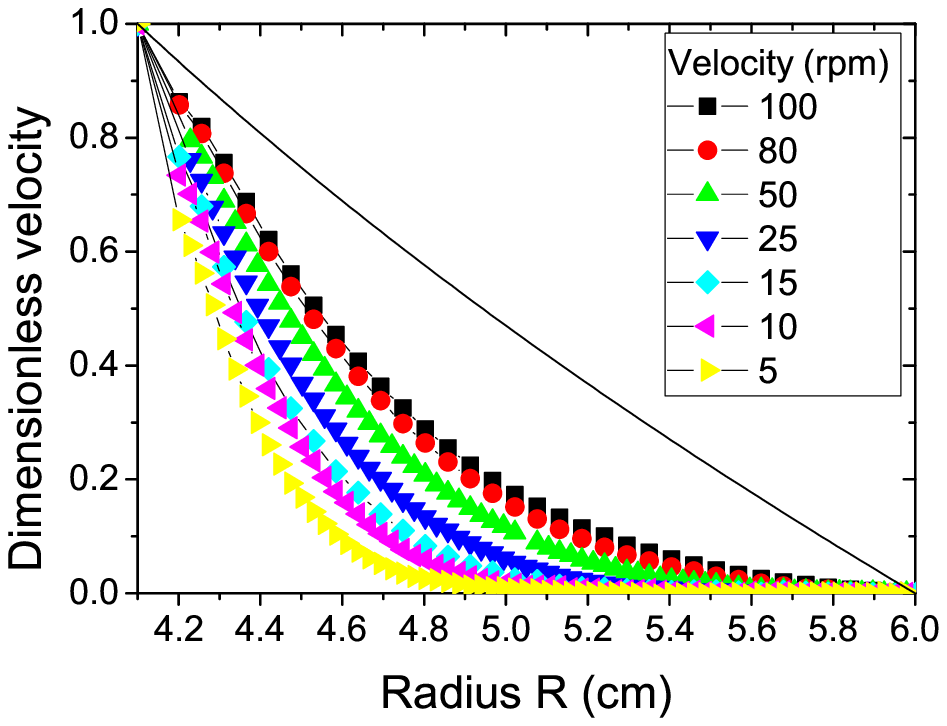}
\includegraphics[width=8cm]{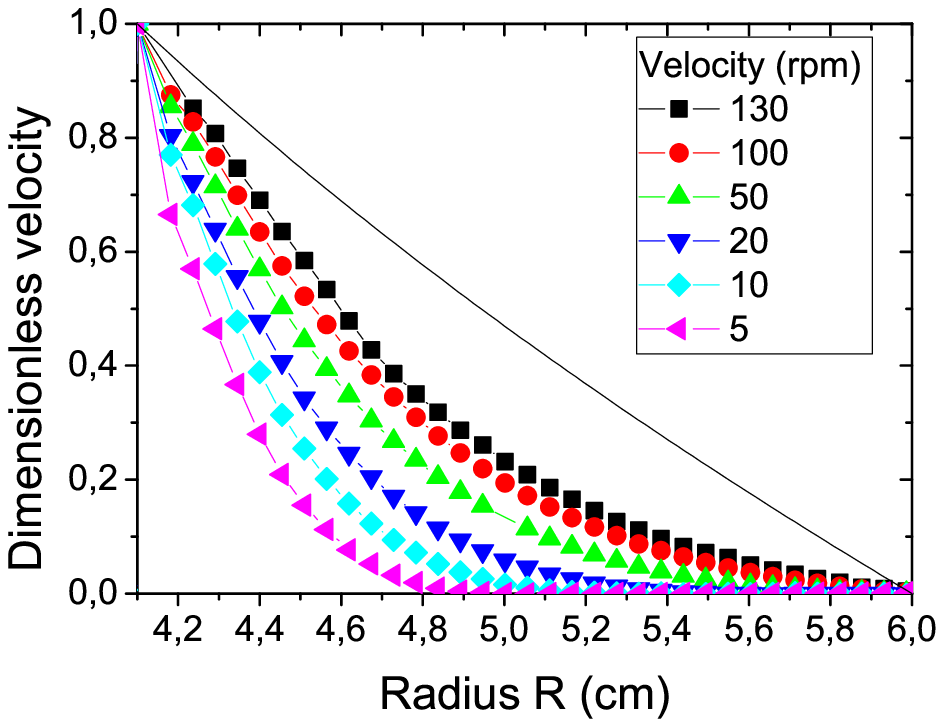}
\includegraphics[width=8cm]{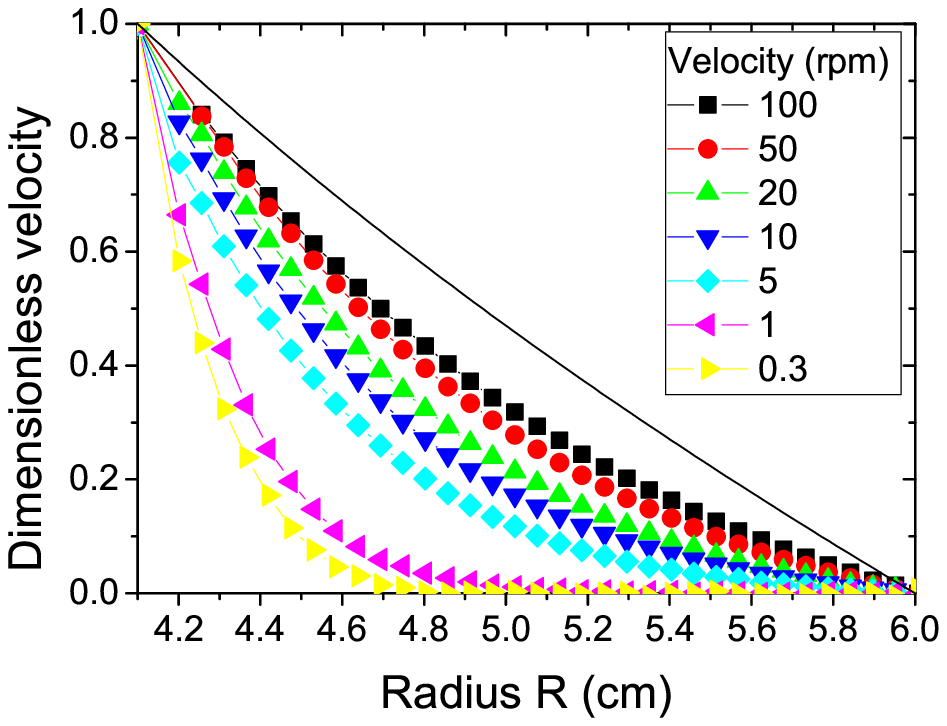}
\includegraphics[width=8cm]{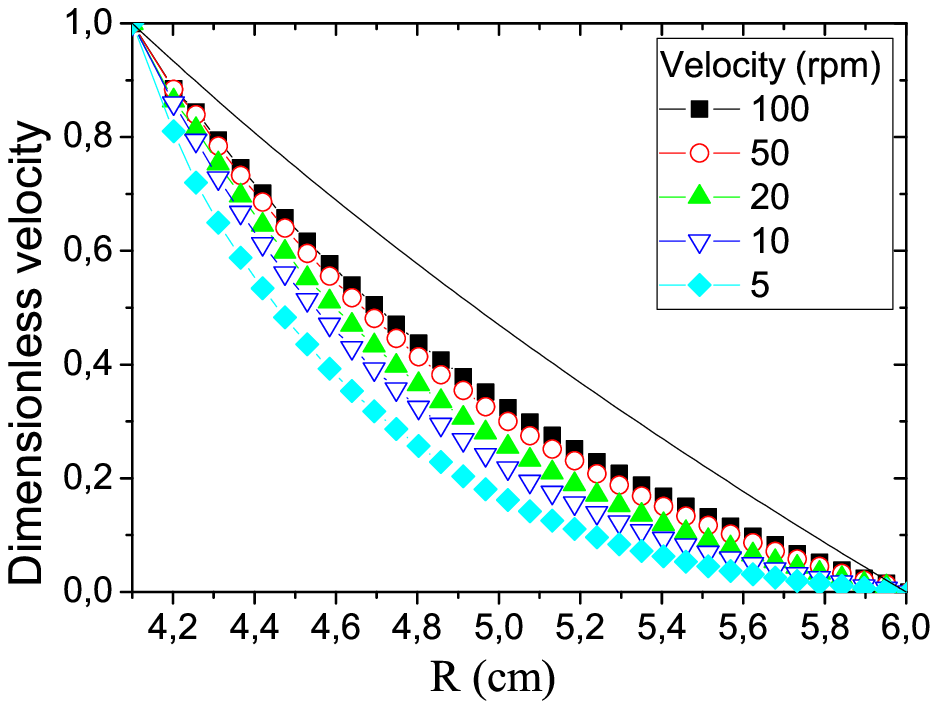}
\includegraphics[width=8cm]{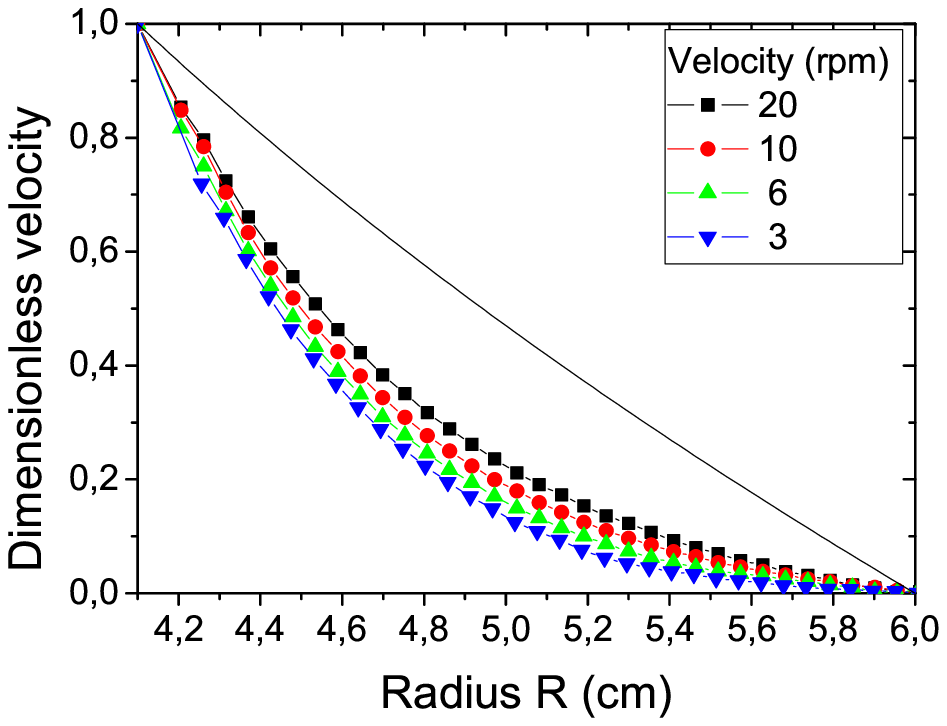}
\caption{(Color online) a) Dimensionless velocity profiles for the
steady flows of a 0.3$\mu$m adhesive emulsion, at various
rotational velocities ranging from 5 to 100rpm; the line is the
theoretical dimensionless velocity profile for a Newtonian fluid.
b) Same plot for a 1$\mu$m non-adhesive emulsion. c) Same plot for
a 6.5$\mu$m adhesive emulsion. d) Same plot for a 6.5$\mu$m
non-adhesive emulsion. e) Same plot for a 40$\mu$m non-adhesive
emulsion.} \label{fig_emul_velocity1}
\end{figure}

In Fig.~\ref{fig_emul_velocity1}, we plot the dimensionless
velocity profiles for the steady flows of all the emulsions for
various rotational velocities ranging from 0.3 to 100rpm.

The 1$\mu$m non adhesive emulsion, the 6.5$\mu$m adhesive and non
adhesive emulsions, and the 40 $\mu$m non adhesive emulsion share
a common behavior. MRI measurements show that the velocity
profiles are curved, and that they occupy only a small fraction of
the gap at low rotational velocities
(Fig.~\ref{fig_emul_velocity1}): in this case, the velocity tends
to zero within the measurement uncertainty at some radius before
the outer cylinder. The fraction of the material that is sheared
increases with the rotational velocity. Beyond a critical velocity
$\Omega_c$, that depends on the emulsion and is of order 3rpm in
the case of the 6.5$\mu$m adhesive emulsion
(Fig.~\ref{fig_emul_velocity1}c), or of the order 50 rpm in the
case of the 1$\mu$m non-adhesive emulsion
(Fig.~\ref{fig_emul_velocity1}b), the whole sample is sheared.
This shear localization is a classical feature of yield stress
fluids flows in Couette geometry where the shear stress is a
decreasing function of the radius: the flow must stop at a radius
$R$ such that the shear stress equals the material yield stress at
this place. Note that this behavior is not evidenced for the non
adhesive 6.5$\mu$m emulsion and for the non adhesive 40$\mu$m
emulsion: this is simply due to the fact that the applied
rotational velocities are too high to probe this behavior.

When all the gap is sheared, the velocity profiles are clearly
different from those of a Newtonian fluid
(Fig.~\ref{fig_emul_velocity1}). The curvature we observe is due
to the use of a wide-gap Couette geometry and is actually typical
of shear-thinning fluid: the shear rate decreases more rapidly
within the gap (i.e when the shear stress decreases) than for a
Newtonian fluid. The shear rate may vary here of a factor 20
within the gap while it would vary of a factor 2.1 for a Newtonian
fluid in this geometry (this factor 2.1 is simply the ratio
between the shear stress at the inner cylinder and the shear
stress at the outer cylinder, see Eq.~\ref{eq_localstress} in
Sec.~\ref{section_torque}). This implies that in the analysis of
macroscopic experiments, one cannot simply compute a mean shear
rate within the gap to characterize the flow.

Finally, note that no data can be recorded close to the inner
cylinder. As a consequence, the expected absence of wall slip (due
to the use of rough surfaces of roughness larger than the droplet
size) cannot be proved at this stage. It can only be observed that
the extrapolation of the dimensionless velocity profiles to a
value of 1 at the inner cylinder is reasonable, and that wall slip
-- if any -- should be of a limited amount of order 5\% or less.
We will show in Sec.~\ref{section_local} that it is possible to
infer the emulsion velocity at the walls from its local
constitutive law, and that wall slip -- if any -- is actually less
than 1\% of the inner cylinder velocity.

\begin{figure}[!h]
\includegraphics[width=8cm]{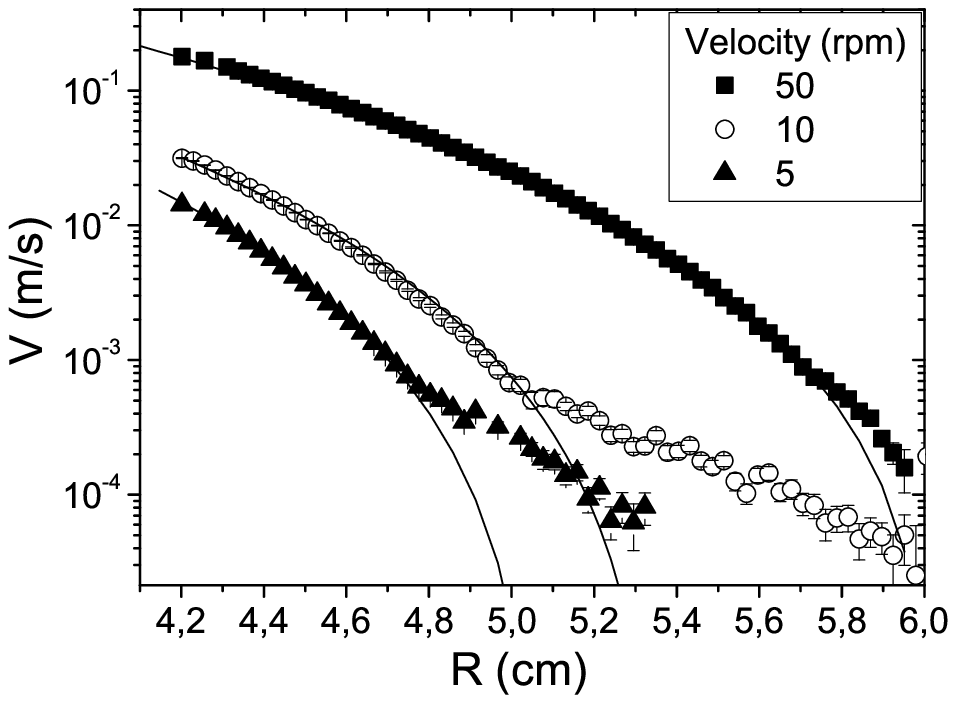}
\includegraphics[width=8cm]{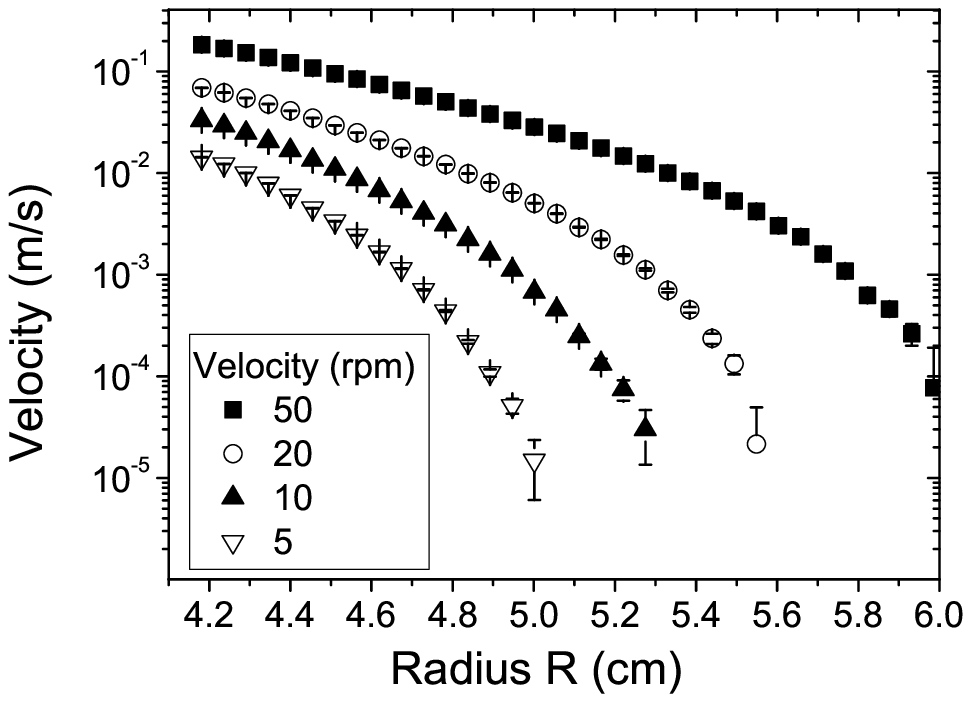}
\includegraphics[width=8cm]{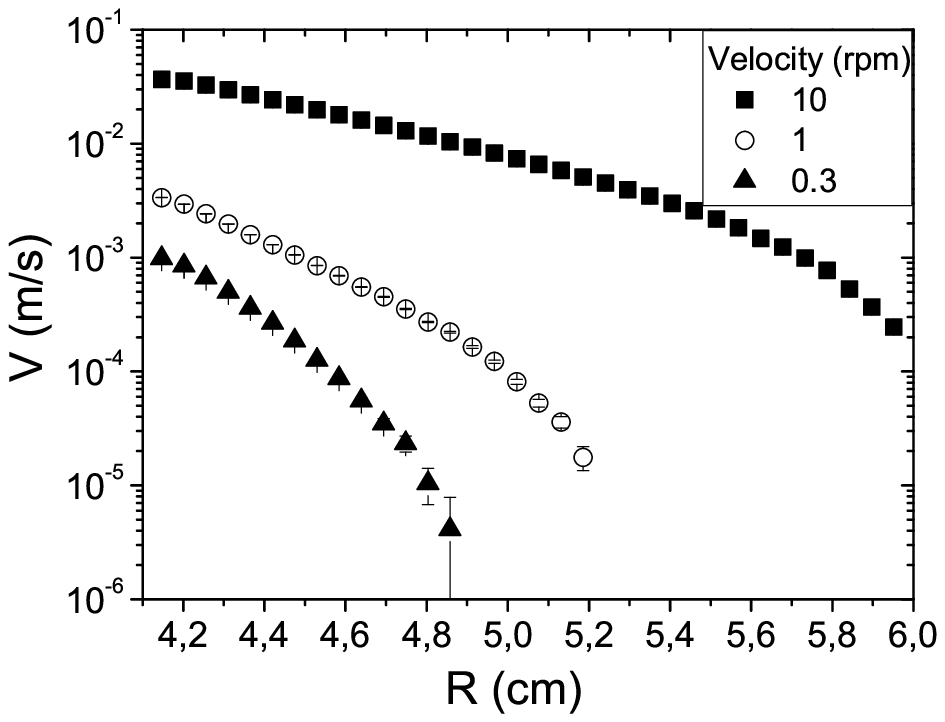}
\includegraphics[width=8cm]{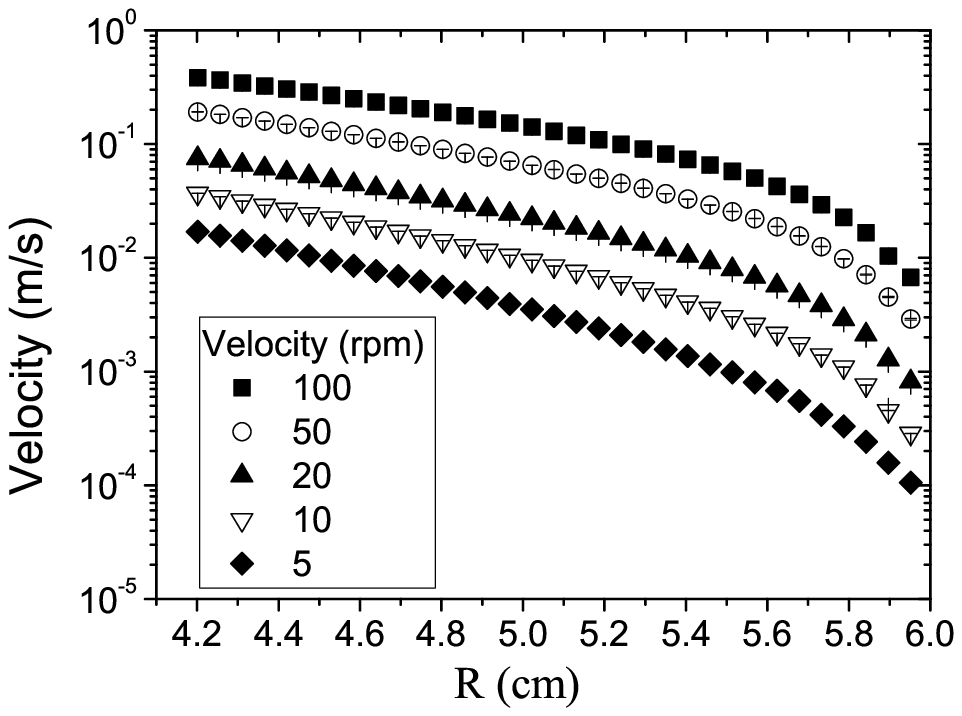}
\includegraphics[width=8cm]{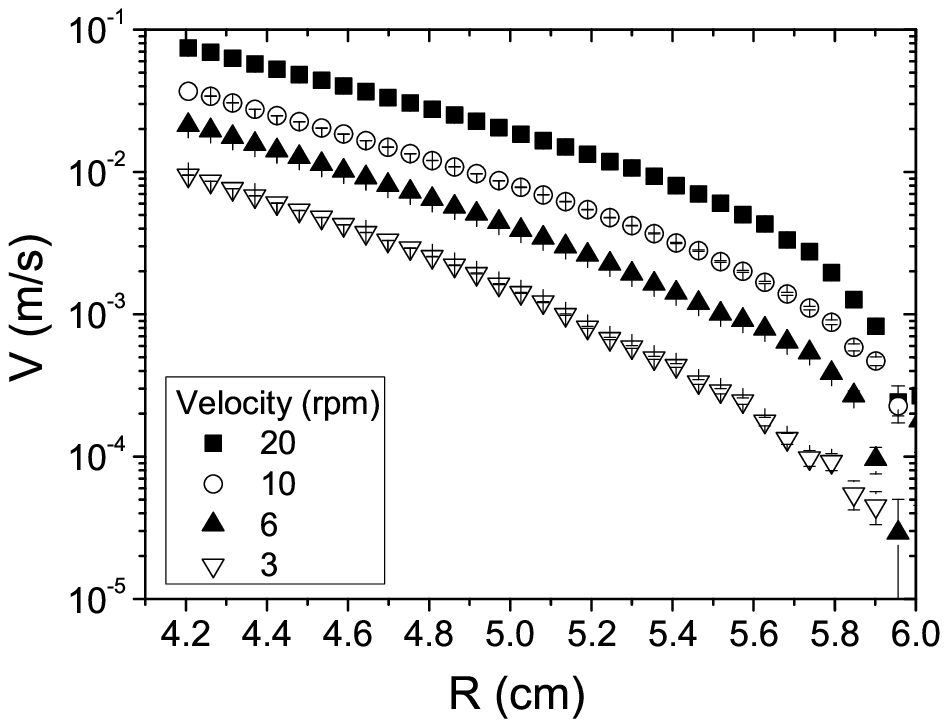}
\caption{Same plots as in Fig.~\ref{fig_emul_velocity1} in log
scale (for readability, not all profiles are plotted). The solid
lines of Fig. \ref{fig_emul_velocity1log}a are the velocity
profiles predicted by a Herschel-Bulkley
law.}\label{fig_emul_velocity1log}
\end{figure}

At first sight, the 0.3$\mu$m adhesive emulsion presents the same
behavior as the other samples. The velocity profiles are curved,
and they occupy only a small fraction of the gap at low rotational
velocities. However, focusing on the low velocities part and thus
very low shear rates part of the profiles, a striking behavior is
evidenced in the semi-logarithmic scale figure (Fig.
\ref{fig_emul_velocity1log}a). A slope break in the profile is
evidenced at a given radius inside the gap. The velocity then
starts to decrease very slowly with the radius. There is then a
large zone of the material in which there remains a very slow
flow. This behavior is reminiscent of the one observed previously
by \citet{becu2006} and by
\citet{Ragouilliaux2006,Ragouilliaux2007} in adhesive emulsions (this behavior will be shortly commented in Sec.~\ref{section_local}).
These slowly varying slow flows occur here for velocities lower
than a few 100$\mu$m.s$^{-1}$. This behavior seems to be a
specific feature of the strongly adhesive 0.3 $\mu$m emulsion. We
measured velocity as low as 10$\mu$m.s$^{-1}$ at the approach of
flow stoppage in the 1$\mu$m non adhesive and 6.5$\mu$m adhesive
emulsion (Fig. \ref{fig_emul_velocity1log}b,c) but did not observe
any slope break in these cases: there is no evidence of any slow
flow in the apparently jammed zones in these emulsions.

In the following, we build the constitutive laws of the emulsions
accounting for their flow properties. To get this information, we
use two different methods. First, focusing on the macroscopic
measurements, we present an experimental procedure allowing us to
get a relationship between local values of the shear stress and
the shear rate by measuring only macroscopic quantities
(Sec.~\ref{section_torque}). Second, we analyze in detail the
velocity profiles and extract a local constitutive law by
differentiating the velocity field to get the local shear rate
(Sec.~\ref{section_local}).

\subsection{Macroscopic characterization}\label{section_torque}

In this section, we present the macroscopic rheometric
measurements (torque vs. rotational velocity). We then show that
it is possible to extract the constitutive law of the material
from a proper analysis of this set of purely macroscopic data only
provided that the material is homogeneous (absence of migration)
and that there is no wall slip, which is what we have ensured.

\begin{figure}[htbp]
\includegraphics[width=8cm]{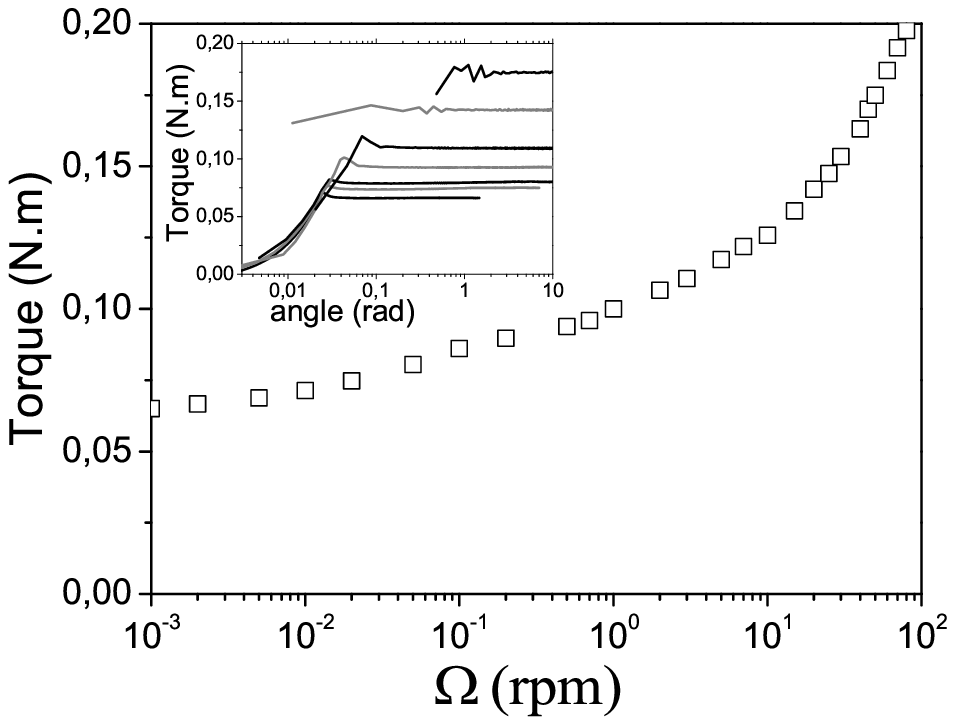}
\includegraphics[width=8cm]{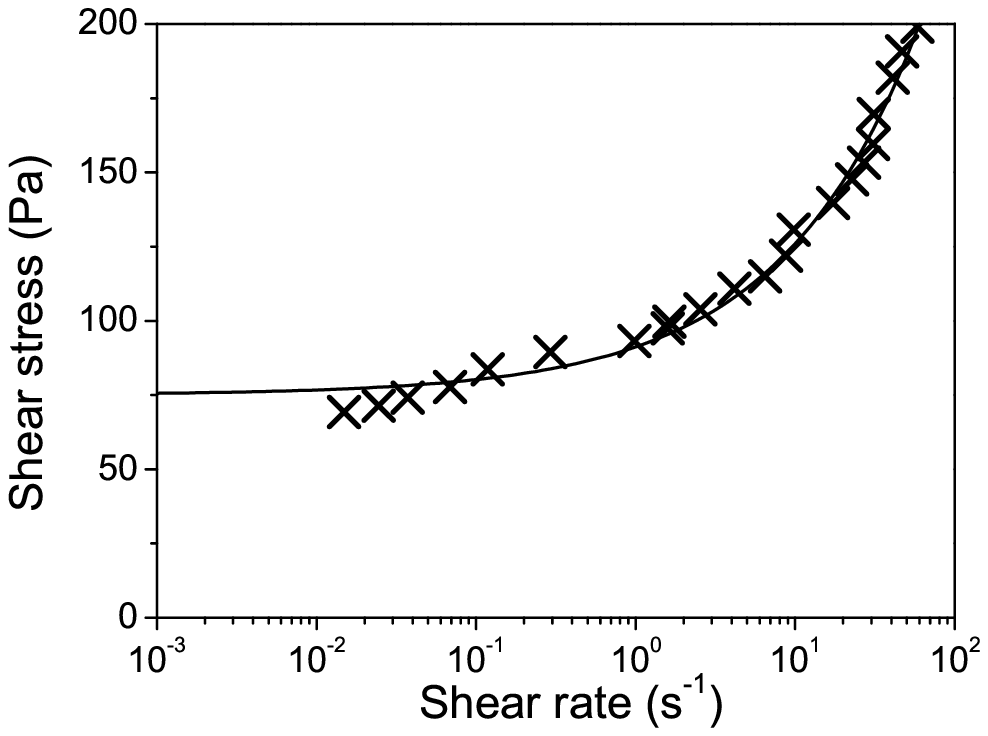}
 \caption{a) Macroscopic
measurements: stationary torque vs. rotational velocity for the
steady flows of a 0.3$\mu$m adhesive emulsion; inset: torque vs.
angular displacement for various rotational velocities ranging
between 0.01 and 50 rpm (bottom to top). b) Constitutive law
inferred from the purely macroscopic measurements thanks to
Eq.\ref{eq_localstress} and Eq.~\ref{eq_localrate}; the solid line
is a Herschel-Bulkley fit $\tau=\tau_c+\eta_{_{HB}}\gdot^n$ of the
data with $\tau_c=77$Pa, $\eta_{_{HB}}=15.4$Pa.s and $n=0.5$.}
\label{fig_emul_rheomacro}
\end{figure}

For each rotational velocity $\Omega$, we have measured the torque
exerted on the inner cylinder vs. time until a stationary state is
reached (see inset of Fig.~\ref{fig_emul_rheomacro}a for the 0.3
$\mu$m adhesive emulsion). We observe that for each rotational
velocity, a stationary state is reached for a strain of a few
unities, consistently with the absence of evolution of the
velocity profiles in time. In Fig.~\ref{fig_emul_rheomacro}a we
plot the stationary torque vs. the rotational velocity for the
steady flows of the 0.3 $\mu$m adhesive emulsion.

We observe that the apparent flow curve of the material is that of
a yield stress shear-thinning fluid. As we have shown in
Sec.~\ref{section_concentration} that the material is homogeneous,
we are allowed to use these macroscopic data to infer the
constitutive law of the material. Because of the strong stress and
shear rate heterogeneities, in particular shear localization,
standard formula relating simply the torque and rotational
velocity values to a mean shear stress and a mean shear rate
cannot be used. However, it is possible to analyze the macroscopic
data to account for these heterogeneities and to obtain the unique
constitutive law consistent with this set of purely macroscopic
data. The method used to analyze the wide gap Couette rheometric
data in order to obtain the constitutive law of the material is
detailed e.g. in \cite{Coussot2005}, we give here the main steps.
First, it has to be noted that the shear stress distribution
$\tau(R)$ within the gap at a radius $R$ is known whatever the
constitutive law of the material; it reads \bea\tau(R)=T/(2\pi H
R^2)\label{eq_localstress}\eea where $T$ is the torque and H is
the height of the Couette geometry. Then, the local shear rate
$\gdot(R)$ is related to the rotational velocity $\Omega$ through
\bea\Omega=\int^{R_o}_{R_i} \gdot(R)/R\ {\rm
d}\!R\label{eq_omega0}\eea where $R_i$ is the inner radius and
$R_o$ is the outer radius. From Eqs. \ref{eq_localstress} and
\ref{eq_omega0}, we see that from the knowledge of the torque $T$
and the rotational velocity $\Omega$, we obtain a non
straightforward relationship between the local shear stress and
the local shear rate. In order to go one step further, Eqs.
\ref{eq_localstress} and \ref{eq_omega0} may be combined into
\bea\Omega=-\frac{1}{2}\int^{\tau(R_o)}_{\tau(R_i)}
\gdot(\tau)/\tau\ {\rm d}\tau\label{eq_omega}\eea Note that the
possibility of shear localization is naturally taken into account
in this equation, where $\gdot(R)=0$ when $\tau(R)<\tau_c$ with
$\tau_c$ the material yield stress. The constitutive law may then
be derived from the whole macroscopic rheometric curve $T(\Omega)$
thanks to the differentiation of Eq. \ref{eq_omega} relative to
$T$: \bea2T\partial\Omega/\partial
T=\gdot(\tau(R_i))-\gdot(\tau(R_o))\label{eq_localrate}\eea
Finally, in order to get the shear rate $\gdot(\tau(R_i))$ at the
inner cylinder as a function of the shear stress at the inner
cylinder, one needs to eliminate the shear rate at the outer
cylinder $\gdot(\tau(R_o))$ from this equation. This can be done
by summing this relationship with a series of successive
decreasing torques (that differ by a factor $R_i^2/R_o^2$ at each
step) chosen to ensure that the shear rate at the outer cylinder
is eliminated in the summation at each step, until a zero shear
rate is reached at the outer cylinder: the summation stops when
the shear stress computed at the outer cylinder at a given step
falls below the yield stress \cite{Coussot2005}. While this
summation is in principle infinite for simple fluids, only a few
steps are needed in the case of yield stress fluids, and the
number of steps is more reduced for wider gaps. In the case of our
Couette geometry, as the ratio between the shear stress at the
inner cylinder and the shear stress at the outer cylinder
($R_i^2/R_o^2$) is equal to 2.1, we see that the data of
Fig.~\ref{fig_emul_rheomacro}a can be easily processed thanks to a
maximum number of two summations of Eq.~\ref{eq_localrate}. From
Eq.~\ref{eq_omega}, note that such analysis is based on the
assumption that there is a local constitutive law characterizing
the material and that there is negligible wall slip.

The constitutive law obtained with this method is depicted in
Fig.~\ref{fig_emul_rheomacro}b. We observe that the behavior of
the material obtained with this procedure is basically well fitted
to a Herschel-Bulkley law.

In Sec.~\ref{section_local} we compare this law to the
constitutive law built from the  local shear rates extracted from
the velocity profiles.

\subsection{Local constitutive laws}\label{section_local}

The constitutive laws of the materials accounting for their
velocity profiles can be built from our experimental data, using
both the velocity profiles and the torque measurements. The stress
distribution $\tau(R)$ within the gap is obtained from the
macroscopic torque measurements thanks to
Eq.~\ref{eq_localstress}. The local shear rate $\gdot(R)$ in the
gap is inferred from the velocity profiles $V(R)$ through
\bea\gdot(R)=V/R-\partial V/\partial R\eea Both measurements
performed at a given radius $R$ for a given rotational velocity
$\Omega$ thus provide a local data point of the constitutive law
$\tau=f(\gdot)$. This analysis provides a fair local measurement
of the constitutive law since we measure the true local shear rate
within the bulk of the material and the shear stress distribution
is known from the momentum balance independently of any
hypothesis; in particular, it is independent of what happens at
the interface so that a possible wall slip does no affect this
analysis. We are finally allowed to combine the data measured at
various radiuses because the materials are homogeneous as shown in
Sec.~\ref{section_concentration}.

\begin{figure}[htbp]
\includegraphics[width=8cm]{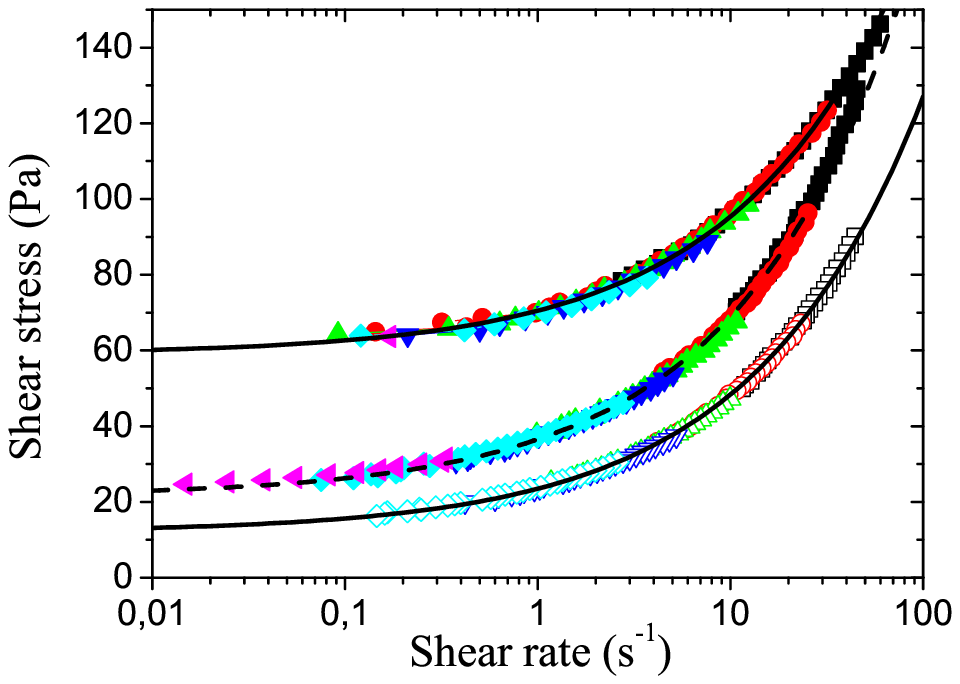}
\includegraphics[width=8cm]{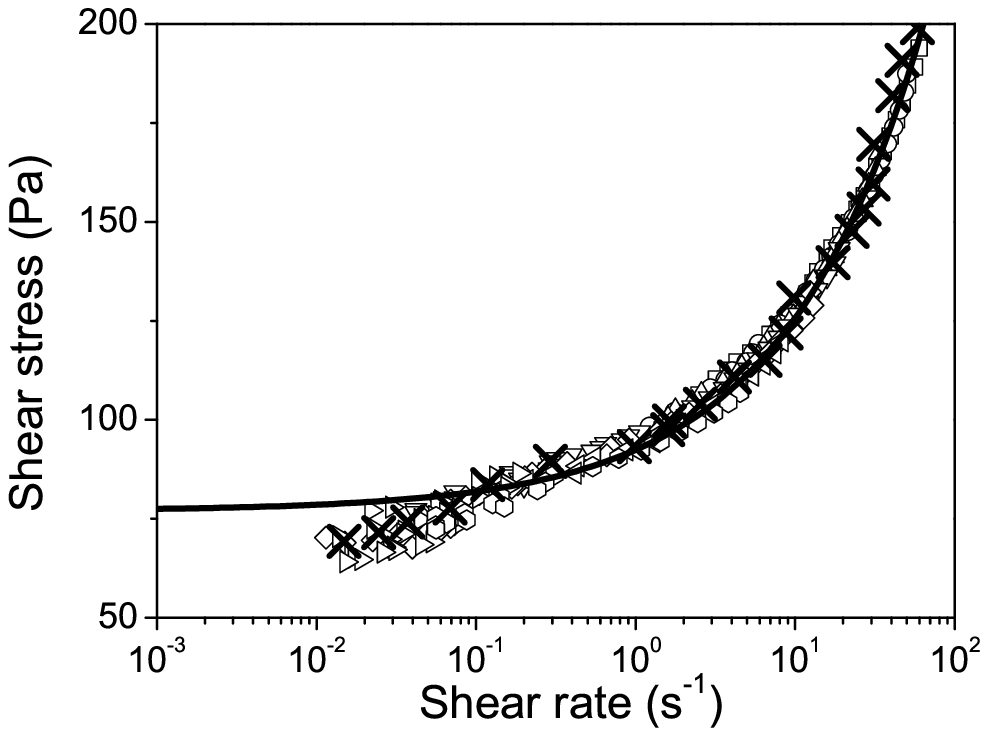}
\caption{(Color online) a) Constitutive law measured locally in
the gap of a Couette cell for the flows of (from top to bottom) a
non-adhesive 1$\mu$m emulsion, a non-adhesive 6.5$\mu$m emulsion
and an adhesive 6.5$\mu$m emulsion; the solid line is a
Herschel-Bulkley fit $\tau=\tau_c+\eta_{_{HB}}\gdot^n$ of the data
with $n=0.5$ for all materials, $\tau_c=59$Pa and
$\eta_{_{HB}}=11.5$Pa.s for the non-adhesive 1$\mu$m emulsion,
$\tau_c=12$Pa and $\eta_{_{HB}}=11.5$Pa.s for the non-adhesive
6.5$\mu$m emulsion, and $\tau_c=21.5$Pa and $\eta_{_{HB}}=15$Pa.s
for the adhesive 6.5$\mu$m emulsion. b)Constitutive law measured
locally in the gap of a Couette cell for the flows of a 0.3$\mu$m
adhesive emulsion (empty symbols); the solid line is a
Herschel-Bulkley fit $\tau=\tau_c+\eta_{_{HB}}\gdot^n$ of the data
with $\tau_c=77$Pa, $\eta_{_{HB}}=15.4$Pa.s and $n=0.5$. The
crosses are the data extracted from the analysis of the purely
macroscopic measurements (see Fig. \ref{fig_emul_rheomacro}b)
}\label{fig_emul_local1}
\end{figure}

The local constitutive law extracted from the experiments
performed at various rotational velocities $\Omega$ on all the
emulsions are plotted in Fig.~\ref{fig_emul_local1} (we could not
analyze the data for the 40$\mu$m emulsion as torque data could
not be measured in this case).

It should first be noted that for each emulsion, all the shear
stress vs. shear rate data fall along a single curve. This means
that for a given material, our different data obtained under
different boundary conditions effectively reflect the behavior of
a single material with a given intrinsic constitutive law in
simple shear. Note also that this is consistent with the absence
of migration: any heterogeneity would have apparently yielded
different constitutive laws for different rotational velocities
\cite{Huang2005,Ovarlez2006}. Our observation of a single local
constitutive law accounting for all flows of the non-adhesive
6.5$\mu$m emulsion (Fig.~\ref{fig_emul_local1}a) contrasts with
the observations of \citet{Goyon2008} on the same system in
microchannel flows of up to 250$\mu$m width: the velocity profiles
observed in the channel cannot be accounted for by a single
constitutive law. This apparent paradox has been solved by
\citet{Goyon2008}: they showed that a rather simple non-local flow
rule accounts for all the velocity profiles. They concluded that
non-local effects should be observable in a zone involving up to
100 particles; beyond this zone, this non-local law reduces to a
local law. This explains why our measurements are unsensitive to
such effects and why we measure a single local constitutive law:
using a wide gap geometry prevents from being sensitive to these
non-local effects. Our observation of a single local constitutive
law accounting for all flows of the adhesive 0.3$\mu$m emulsion
(Fig.~\ref{fig_emul_local1}b) also contrasts with the observations
of \citet{becu2006}. On exactly the same 0.3$\mu$m adhesive
emulsion sheared in a thin gap Couette geometry, \citet{becu2006}
were actually unable to fit all their velocity profiles with a
single constitutive law. In the \citet{becu2006} study, the gap
being 1mm for 0.3$\mu$m droplets, the non-local effects evidenced
by \citet{Goyon2008} cannot be the reason for the apparent absence
of a single constitutive law accounting for all flows. An
important point should be noted that may explain the differences
between both studies of the same material: the experimental
procedures are slightly different, and the time evolutions of the
emulsion properties are very different. In our experiments, the
emulsion was first presheared with a mixer and then at 100 rpm for
5 minutes in a Couette cell with rough surfaces. The stationary
velocity profiles then measured at various rotational velocities
were found to develop within a few seconds and to remain stable
for hours. In the experimental procedure of \citet{becu2006}, each
experiment is conducted on a fresh sample which is directly
sheared at the studied shear rate during at least 3 hours. A key
point may then be that \citet{becu2006} found their system to
evolve for at least two hours, in contrast with our observations:
whatever the rotational velocity, they found the apparent
viscosity of the system to decrease slowly in time. This would
mean that their material was in an initially structured state, and
was destructured by shear, whatever the magnitude of the shear
rate. Then, while the authors claim that the velocity profiles no
longer change significantly after 2 h, one may think that some
slow relaxation still occurs and that the steady state is not
really reached. This would mean that the measurements performed at
different rotational velocities have been performed on different
structural states of the emulsion (in contrast with our
measurements), naturally leading to different constitutive laws
for the different velocities studied. This explanation based on
the influence of the initial state of the material would be
consistent with the observations of \citet{Ragouilliaux2007} who
showed that simple emulsions whose flow properties show no
apparent thixotropic effects when first presheared at high shear
rate are actually subject to significant aging at rest: this imply
in particular that such systems have to be strongly presheared
before any study. This preshear was performed in our study, it was
not in the \citet{becu2006} study. Two other points may be noted:
first, although the composition of the two emulsions is the same,
the samples are different and chemical impurities may play an
important role in the adhesion process; second, the Couette cell
used by \citet{becu2006} is smooth and huge wall slip is
evidenced.

Focusing on the non adhesive 1$\mu$m, the adhesive 6.5 $\mu$m, and
the non adhesive 6.5 $\mu$m emulsions, we observe that the
behavior of all materials is well fitted to a Herschel-Bulkley
behavior $\tau=\tau_c+\eta_{_{HB}}\gdot^n$ of index $n=0.5$ in all
the range of measured shear rates; this is consistent with
previous observations of the behavior of dense emulsions
\cite{Mason1996a,Cloitre}. Coming back to the 0.3$\mu$m adhesive
emulsion, at moderate and high shear rates (above 0.1s$^{-1}$),
the behavior is the same as the one encountered for the large
droplets emulsions: the constitutive law is well fitted to an
Herschel-Bulkley law on index n=0.5.

At low shear rate (in a 0.01-0.1s$^{-1}$ range), the behavior of
the 0.3$\mu$m adhesive emulsion differs from the previous ones.
All the points still fall on the same curve but a slight slope
break is noticed in the flow curve. While a stress plateau should
be reached for a simple yield stress fluid, the shear stress
continues decreasing when the shear rate decreases. This is also
observed on the law inferred from the macroscopic measurement. As
far as we can say, this behavior is specific to the 0.3$\mu$m
adhesive emulsion: we took care of measuring the local behavior at
low shear rates in the 6.5$\mu$m adhesive emulsion but observed no
slope break for shear rates as low as 10$^{-2}$s$^{-1}$. A
probable reason for the differences observed between these systems
is that the 0.3 micron emulsion is more adhesive than the others,
in particular because of their larger surface/volume ratio. This
low shear rate behavior corresponds to the slow flows observed
below a first apparent yield stress in
Fig.~\ref{fig_emul_velocity1log}a. At low rotational velocity,
when there is apparently shear localization, the velocity profiles
are well fitted to the Herschel-Bulkley law predictions only for
velocities larger than a few 100$\mu$m.s$^{-1}$. For lower
velocities, at the approach of the radius where the yield stress
should be reached and the velocity should tend to zero, the
velocity starts to decrease very slowly with the radius; on the
other hand, at high rotational velocity, when all the gap is
sheared, the velocity profile is very well fitted to the
Herschel-Bulkley law prediction. The analysis of this behavior is
difficult. One may think that the emulsion flows homogeneously and
really follows a single continuous local constitutive law that
corresponds to a Herschel-Bulkley equation at high shear rate and
differs from this equation at low shear rate. One may also suggest
that the law at low shear rates does not correspond to steady
flows. Indeed, it has to be noted that the measurements of the
very low velocities in Fig.~\ref{fig_emul_velocity1log}a are
averages over 1min: we cannot know if they correspond to
steady-flows (the measurements may e.g. reflect a stick-slip
behavior). These flows may thus reflect the existence of a jammed
phase perturbated by some unsteady local rearrangements. In this
case, this behavior would be reminiscent of shear banding, which
has already been observed for strongly adhesive emulsions
\cite{Ragouilliaux2007} and is explained by a competition between
structuration due to the adhesion process and destructuration by
shear \cite{Ragouilliaux2007,Coussot2005}.

\subsubsection*{Comparison between local measurements and purely macroscopic measurements}\label{section_macro_vs_local}

The local measurements obtained for the 0.3$\mu$m adhesive
emulsion are compared in Fig.\ref{fig_emul_local1}a with the law
inferred from the purely macroscopic measurements (see
Sec.\ref{section_torque}). We observe that both laws are very well
matched. In particular, the macroscopic measurements also allow us
to evidence the slope break in the constitutive law at low shear
rate. This validate the use of a wide gap Couette geometry as a
tool to obtain the constitutive law of dense emulsions from purely
macroscopic rheometrical measurements, provided that a proper
analysis of the macroscopic data is performed. We recall that this
agreement is due to four important features (i) we have shown that
all dense emulsions seem to be free from migration and thus remain
homogeneous during the experiments, (ii) the use of a wide gap
prevents from the nonlocal effects observed by \citet{Goyon2008},
(iii) the material is initially destructured and a steady state is
studied (this may have not be the case in the \citet{becu2006}
study), (iv) wall slip has to be negligible to allow for a proper
analysis of the macroscopic data: this is ensured here by the use
of rough walls of roughness larger than the droplet size, as shown
below.

\subsubsection*{Wall slip}

As pointed out in Sec.~\ref{section_velocity}, our velocity
measurement method does not provide the velocity at the walls: the
first reliable data is obtained at around 0.5 to 1mm from the
inner cylinder. However, the local measurements of the
constitutive law we have performed make it possible to study in
more detail a possible wall slip effect. For a given rotational
velocity of the inner cylinder, the velocity of the emulsion
expected at the walls can actually be computed thanks to the
knowledge of the shear rate in the emulsion corresponding to the
shear stress value at the wall. The same shear stress level was
actually reached somewhere in the bulk of the material for a
higher rotational velocity of the inner cylinder, and the shear
rate corresponding to this shear stress was then measured locally
thanks to the velocity profile measurement. The velocity profile
expected in the zone near the inner cylinder can then be
reconstructed from the velocity profile measured at some distance
from the walls and from the expected shear rate near the wall. Of
course, this analysis is based on the hypothesis that the material
is the same as in the bulk in this 0.5 to 1mm zone near the walls,
which may be no more true in a zone where there are nonlocal
effects \cite{Goyon2008}. Note also that this method can be used
only when the reproducibility of the experiments is very good
(allowing data measured at a given rotational velocity to be used
to predict what happens at another velocity).

\begin{figure}[htbp]
\includegraphics[width=8cm]{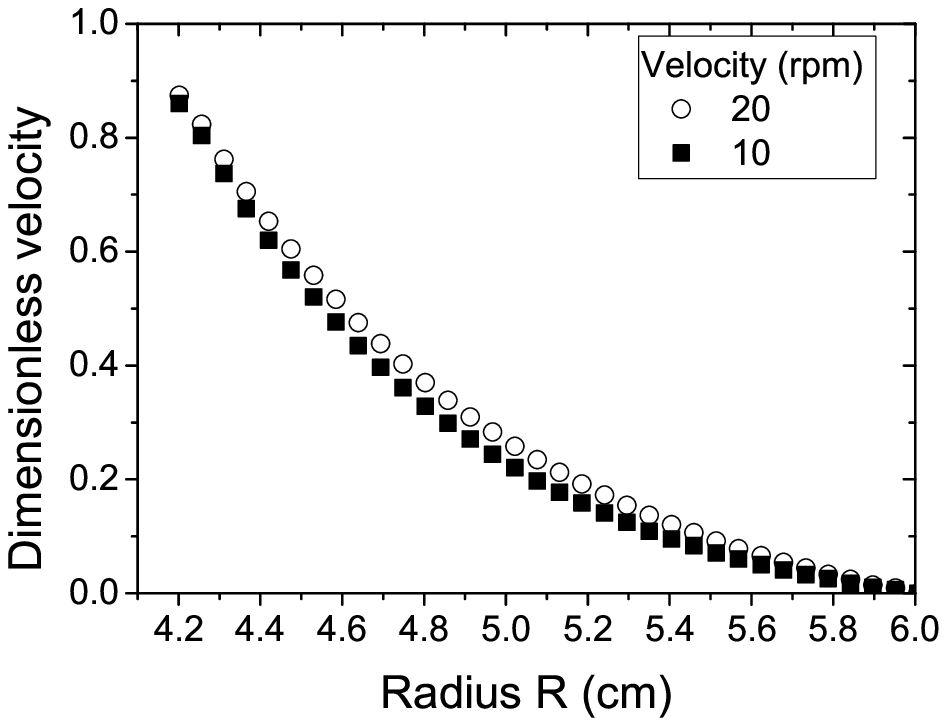}
\includegraphics[width=8cm]{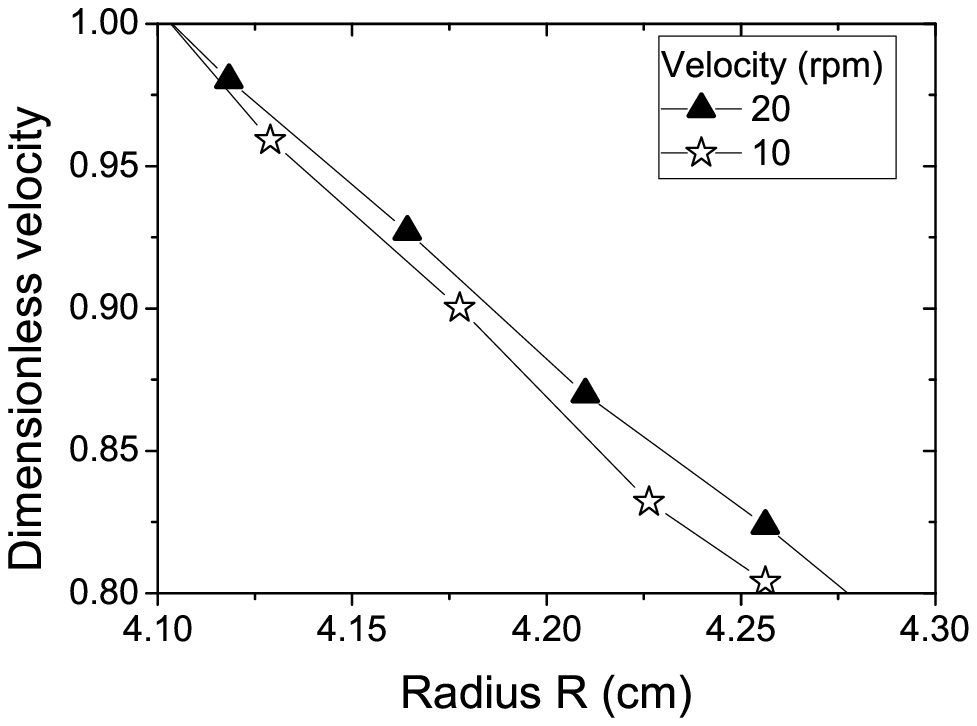}
\caption{a) Dimensionless velocity profiles for the steady flows
of a 6.5$\mu$m adhesive emulsion, for a 10 and a 20 rpm rotational
velocity. b) Dimensionless velocity profiles expected near the
walls at 10 and 20 rpm; the velocities are computed thanks to the
local shear rate measurements performed at 50
rpm.}\label{fig_emul_velocity2_slippage}
\end{figure}

An example of this method applied to the flows of the 6.5$\mu$m
adhesive emulsion is depicted in
Fig.~\ref{fig_emul_velocity2_slippage}.

We observe in Fig.~\ref{fig_emul_velocity2_slippage}b that the
dimensionless velocity profiles expected near the walls have a
value that is very close to 1. This analysis was performed on
several velocity profiles, and in all cases the dimensionless
velocity values predicted at the inner cylinder were found to be
equal to 1 within 1\%. We finally conclude that, thanks to the use
of rough surfaces of roughness larger than the droplet size, there
is probably no wall slip in our experiments; if any, it is of
order 1\% or less. This feature shows that the macroscopic
measurements performed with a wide gap Couette rheometer with
rough walls can be trusted: this explains the very good agreement
observed in Fig.~\ref{fig_emul_local1}a between the constitutive
law measured locally and the constitutive law inferred from the
purely macroscopic measurements.

\section{Conclusion}

We have studied the flows of several dense emulsions in a wide gap
Couette geometry. We have coupled macroscopic rheometric
experiments and local velocity and concentration measurements
through MRI techniques. The method devoted to measure the local
droplet concentration was developed specifically for this study.
In contrast with dense suspensions of rigid particles where very
fast migration occurs under shear, we showed for the first time
that no migration takes place in dense emulsions even for strain
as large as 100000 in our systems. This may imply that another
mechanism is involved in dense emulsions than in dense
suspensions. The homogeneity of our materials under shear allows
to infer their constitutive law from purely macroscopic
measurements. This constitutive law is consistent with the one
inferred from the velocity profiles. This contrasts with previous
results obtained by \citet{Goyon2008} within a microchannel where
nonlocal finite size effects are likely to have prevented from
obtaining a constitutive law. It also differs from the results of
\citet{becu2006}, where the loading procedure probably prevented
the authors to reach a steady state. We thus suggest that properly
analyzed purely macroscopic measurements in a wide gap Couette
geometry can be used as a tool to study dense emulsions. All
behaviors we observed are basically consistent with
Herschel-Bulkley laws of index 0.5. However, we also evidence the
existence of discrepancies with this law at the approach of the
yield stress due to slow shear flows in the case of strongly
adhesive emulsions, whose physical origin is unclear. Finally, we
have shown that there is probably no wall slip.

\acknowledgments We thank S\'ebastien Manneville for fruitful
discussions. Financial support of Rhodia and R\'egion Aquitaine is
acknowledged.

\end{document}